\documentclass[prl,showpacs,nofootinbib,superscriptaddress,twocolumn,preprintnumbers]{revtex4}
\usepackage[mathscr]{eucal}
\usepackage[latin1]{inputenc}
\usepackage{amsmath}
\usepackage{amsfonts}
\usepackage{amssymb}
\usepackage{pictex}
\def\bb{\tt {b}}

\begin{document}
\title{Reply to `Comment on ``Quantum Bounce and Cosmic Recall'''}
\author{Alejandro Corichi}
\affiliation{
Instituto de Matem\'aticas, Unidad Morelia, \\
Universidad Nacional Aut\'onoma de M\'exico, UNAM-Campus Morelia,
\\
A. Postal 61-3, Morelia, Michoac\'an 58090,
Mexico}
\author{Parampreet Singh}
\affiliation{
Perimeter Institute for Theoretical Physics,\\
 31 Caroline Street
North, Waterloo, Ontario N2L 2Y5, Canada}

\begin{abstract}
A recent Comment \cite{MB-Co} on a Letter by the authors
\cite{CS-PRL} is shown to arise from an incorrect understanding of
the issues at hand and of our analysis. The conclusions of
Bojowald's Comment are shown to add little to our work, to be
irrelevant at best, and are further shown to be in contradiction
with his own claims in the literature.
\end{abstract}

\maketitle

\noindent 
In the Letter \cite{CS-PRL} an exactly solvable model within Loop
Quantum Cosmology is considered, for which the genericity of a
quantum bounce replacing the big bang and a supremum on energy
density was established in \cite{SLQC}. These studies show the
robustness of the results first obtained in \cite{APS}. The
question posed in the Letter is: Could a semiclassical state at
late times, have evolved from an arbitrary state at early times
before the bounce? For that the standard definition of a
semiclassical state peaked on a classical trajectory was used,
requiring (i) the expectation values for a complete set of
observables are close to their classical values and (ii) the
quantum fluctuations of the observables are much smaller than
their mean values. Our analysis does not require coherence, a much
stronger assumption.

In \cite{CS-PRL} it was first shown that for a large class of
states,   the fluctuations are perfectly symmetric across the
bounce. Second, for generic states a bound was proven on the
relative fluctuation of volume at early times, assuming that the
state was semiclassical at late times. The bound constrains the
relative fluctuation to be very small compared to unity for
realistic universes. Is the state before the bounce semiclassical?
Yes. Is it possible that the relative fluctuations can be very
different on one side and the other? Yes. Are these two statements
contradictory? No. As discussed in \cite{CS-PRL}, the absolute
fluctuations can change
significantly  across the bounce, allowing for 
the relative fluctuations to change considerably when compared at
the same volume across  the bounce (expectation values of volume
are symmetric \cite{SLQC}). However as strongly cautioned in the
Letter, this does not affect the semiclassical properties of the
state, since the relative fluctuations remain much smaller than
unity.

In the Comment \cite{MB-Co}, the author points out and repeats
many of the statements of the previous paragraph, using the
numerical values that were employed in an example in
\cite{CS-PRL}. For instance, by ignoring that our bound is only an
upper bound, it is pointed out that it  would be consistent with a
relative dispersion in volume of $10^{-28}$ for early times, when
the dispersion, at late times is assumed to be of the order of
$10^{-56}$. Author then comments that our bound is weak and that
in the example the state is assumed to be minimum uncertainty on
one side but does not retain this feature on the other side of the
bounce. Three remarks are in order.

(i) The bound of \cite{CS-PRL}, as was strongly emphasized there,
is  weak since it makes almost no assumption about the initial
state (other than assuming semiclassicality), and is valid for a
large class of states which may not be coherent. The bound would
not be stringent for certain aspects of the state. In fact it was
known that the change of relative dispersions is much smaller than
the bound for reasonable states \cite{APS}. The importance of our
bound is that it proves that the state remains semiclassical,
ruling out claims such as: ``It is practically impossible to draw
conclusions about fluctuations of the Universe before the Big
Bang" \cite{nature}. (ii) In the particular case mentioned in
\cite{MB-Co}, a state with a relative dispersion of $10^{-28}$
{\it is} strongly peaked and certainly semiclassical by any
standard convention. Even when $10^{-28}$ is large relative to
$10^{-56}$, it {\it is} a very small number compared to unity.
(iii) As emphasized, our analysis neither requires minimum
uncertainty states nor the ones such that $(\Delta
\bb/\bb)$/$(\Delta V_{\phi}/\langle \hat V_{\phi} \rangle) \sim
O(1)$. Even if the latter is $10^{20}$ and the initial state was
off from minimum uncertainty by a similar factor, one obtains $D <
10^{-36}$ where $D$ is change in relative fluctuation of volume
\cite{CS-PRL}. The state retains semiclassicality across the
bounce.

In \cite{MB-Co} it is stated that the analysis of \cite{CS-PRL} is
``intrinsically inconsistent''. As we show above, we could not
disagree more. Further, the analysis of \cite{harmonic} based on
dynamical coherent states, when corrected and adjusted to the
system under consideration (the unique  consistent loop
quantization \cite{CS-unique}), shows that the relative change of
relative dispersion in volume, for dynamical coherent states, is
at most of the order of ``20'' across the bounce. For a 1 MPc
universe, this implies $D < 10^{-112}$. These results show the
mutual consistency of the corrected analysis of \cite{harmonic}
with our results, pointing to a recall for the semiclassical
properties of the universe across the bounce and falsifying the
claims of \cite{nature}.

\noindent There is no cosmic forgetfulness.



\begin{thebibliography}{99}

\bibitem{MB-Co} M.~Bojowald,
  ``Comment on 'Quantum bounce and cosmic recall' [arXiv:0710.4543],''
  Phys.\ Rev.\ Lett.\  {\bf 101}, 209001 (2008)
  {\tt arXiv:0811.2790 [gr-qc]}.


\bibitem{CS-PRL} A.~Corichi and P.~Singh,
 ``Quantum bounce and cosmic recall,''
 Phys. Rev. Lett. {\bf 100}, 161302 (2008).
 {\tt arXiv:0710.4543 [gr-qc]}.


\bibitem{SLQC} A.~Ashtekar, A.~Corichi and P.~Singh,
``Robustness of key features of loop quantum cosmology,'' Phys.\
Rev.\ D {\bf 77}, 024046 (2008).
{\tt arXiv:0710.3565 [gr-qc]}.


\bibitem{APS} A.~Ashtekar, T.~Pawlowski and P.~Singh,
  ``Quantum nature of the big bang: Improved dynamics,''
  Phys.Rev.Lett. 96 (2006) 141301, Phys.\ Rev.\ D {\bf 74}, 084003 (2006).
  {\tt arXiv:gr-qc/0607039}.


\bibitem{nature}
M. Bojowald,  ``What happened before the Big Bang?'', Nature
Physics  {\bf 3}, 523 - 525 (01 Aug 2007).


\bibitem{harmonic}   M.~Bojowald,
 ``Harmonic cosmology: How much can we know about a
universe before the big  bang?,'' {\tt arXiv:0710.4919 [gr-qc]}.

\bibitem{CS-unique} A.~Corichi and P.~Singh,
  ``Is loop quantization in cosmology unique?,''
  Phys.\ Rev.\  D {\bf 78}, 024034 (2008).
 {\tt arXiv:0805.0136 [gr-qc]}.





\end{thebibliography}
\end{document}